\documentclass[apj]{emulateapj}
\usepackage{amssymb,amsmath}
\usepackage{graphicx}
\usepackage{wrapfig}
\usepackage[all]{xy}
\usepackage{hyperref}
\newcommand{\nc}{\newcommand}
\nc{\beq}{\begin{equation}}
\nc{\eeq}{\end{equation}}
\nc{\bea}{\begin{eqnarray}}
\nc{\eea}{\end{eqnarray}}
\nc{\n}{\nonumber \\}
\nc{\mt}{\langle\tau\rangle}
\nc{\rt}{\tau_{\rm RMS}}
\nc{\corr}{\langle \delta f \, \delta g \rangle}
\newcommand{\kb}{k_{\mathrm{o}}}
\newcommand{\bo}{b_{\mathrm{o}}}
\newcommand{\al}{\alpha}

\begin{document}  

\title{Reionization on Large Scales II: Detecting Patchy Reionization through Cross Correlation of the Cosmic Microwave Background}
\author{A. Natarajan \altaffilmark{1}, N. Battaglia \altaffilmark{1}, H. Trac \altaffilmark{1}, U.-L. Pen \altaffilmark{2}, A. Loeb \altaffilmark{3}}

\altaffiltext{1}{McWilliams Center for Cosmology, Carnegie Mellon University, Department of Physics, 5000 Forbes Ave., Pittsburgh PA 15213, USA}

\altaffiltext{2}{Canadian Institute for Theoretical Astrophysics, University of Toronto, 60 St. George Street, Toronto, ON M5S 3H8, Canada}

\altaffiltext{3}{Institute for Theory \& Computation, Harvard University, 60 Garden Street, Cambridge, MA 02138, USA}

\begin{abstract}
We investigate the effect of patchy reionization on the cosmic microwave background temperature. An anisotropic optical depth $\tau (\hat n)$ alters the $TT$ power spectrum on small scales $l > 2000$. We make use of the correlation between the matter density and the reionization redshift fields to construct full sky maps of $\tau(\hat n)$. Patchy reionization transfers CMB power from large scales to small scales, resulting in a non-zero cross correlation between large and small angular scales. We show that the patchy $\tau$ correlator is sensitive to small root mean square values $\rt \sim$ 0.003 seen in our maps. We include frequency independent secondaries such as CMB lensing and kinetic Sunyaev-Zeldovich (kSZ) terms, and show that patchy $\tau$ may still be detected at high significance. Reionization models that predict different values of $\rt$ may be distinguished even for the same mean value $\mt$. It is more difficult to detect patchy $\tau$ in the presence of  larger secondaries such as the thermal Sunyaev-Zeldovich (tSZ), radio background, and the cosmic infrared background.  In this case, we show that patchy $\tau$ may be detected if these frequency-dependent secondaries are minimized to $\lesssim 5 \, \mu$K (root mean square) by means of a multi-frequency analysis. We show that the patchy $\tau$ correlator provides information that is complementary to what may be obtained from the polarization and the kSZ power spectra.
\end{abstract}

\keywords{Cosmology: theory, reionization, cosmic background radiation.}

\section{Introduction}

Studies of  high redshift  quasar spectra obtained by the Sloan Digital Sky Survey have found near complete absorption of light blueward of the Ly$\alpha$ line for a quasar at redshift $z=6.28$ \citep{fan_etal2001}, i.e. the first detection of the Gunn-Peterson \citep{gunn_peterson1965} trough. Spectroscopy using the Keck telescope \citep{becker_etal2001} and the Very Large Telescope (VLT)  \citep{pentericci_etal2002}  confirmed the earlier results.  These observations, and the non-detection of the Gunn-Peterson trough for $z < 5.5$ \citep{becker_etal2001} imply that the Universe is highly ionized today \citep{fan_etal2002}.

Evidence for reionization at the $5.5\sigma$ level was obtained by the Wilkinson Microwave Anisotropy Probe (WMAP) measurement of the CMB $EE$ polarization power spectrum \citep{larson_etal2011}, implying a reionization optical depth $\tau = 0.089 \pm 0.014$ \citep{wmap9a, wmap9b}. If interpreted in terms of a single step, sudden reionization model, the epoch of reionization $\approx10.5 \pm 1.2$ \citep{larson_etal2011}. The WMAP data is also consistent with an extended reionization scenario \citep{dunkley_etal2009}.

Reionization began at a redshift $z \sim 20-30$ when the first stars were formed. Later, Population II stars, star forming galaxies, and active galactic nuclei completed the process \citep{tumlinson_shull, loeb_barkana2001, barkana_loeb2001, wyithe_loeb2003, ciardi_etal2003, sokasian_etal2003}.  Due to the very large photoionization cross section at energies $\gtrsim 13.6$ eV, ultraviolet (UV) photons are absorbed by gas in the immediate vicinity of the sources, forming ``bubbles'' of ionized Hydrogen. This leads to large spatial fluctuations in the ionized fraction, and reionization is said to be \emph{patchy}. The bubbles grow and eventually merge, resulting in a uniformly reionized Universe at $z \sim 6$. 

 If reionization is patchy,  scattering of CMB photons varies with the line of sight, introducing secondary anisotropies. \citet{zahn_etal2005} and \citet{mcquinn_etal2006} used analytical models and numerical simulations to study patchy reionization. \citet{weller1999} and \citet{liu_etal2001} computed the modification to the $EE$ polarization caused by inhomogeneous reionization, and found the effect to be small, though potentially detectable by future observations. \citet{hu2000} showed that inhomogeneities in the free electron density could generate $B$ mode polarization through Thomson scattering, while \citet{dore2007} found that the polarization power due to patchy $\tau$ has a unique signature. \citet{vera_etal} recently discussed patchy screening of the CMB by inhomogeneous reionization using off diagonal $TB$ and $TT$ correlations in the WMAP-7 temperature and polarization data.
 
 It was shown by \citet{dvorkin_etal2009} that in addition to $B$ modes being created by inhomogeneous Thomson scattering, patchy screening of the primary $E$ mode leads to $B$ mode polarization. \citet{mortonson_hu2010} used the South Pole Telescope (SPT) limits on secondary anisotropies at $l=3000$ to infer that fluctuations in the optical depth are utmost a few percent of the mean value $\mt$.  Recent observations by the SPT collaboration \citep{spt1} have placed an upper limit on the patchy kinetic Sunyaev-Zel'dovich (kSZ) power at $l=3000$ to be $D^{\rm patchy}_{3000} < 4.9 \, \mu$K$^2$ at the 95\% confidence level when the degree of angular correlation between the thermal Sunyaev-Zel'dovich (tSZ) and the cosmic infrared background (CIB) is allowed to vary. The SPT results \citep{spt1} imply that reionization ended at $z > 5.8$ at 95\% confidence (accounting for the tSZ-CIB correlation), in good agreement with quasar observations.

\citet{dvorkin_smith2009} described a quadratic estimator for reconstructing patchy reionization from the CMB, by computing cross correlations of the $T$, $E$, and $B$ fields, as a function of multipole $l$. They found that nearly all the signal-to-noise comes from the $EB$ cross correlation. This cross power is nonzero because   modulation of the large scale reionization $E$-mode by the smaller scale $\tau$ fluctuations generates $B$-modes, and also because patchy screening of the CMB results in $B$-modes.  While this patchy $\tau$ estimator is optimal, it is unlikely that ongoing experiments will measure the $B$-mode with high accuracy.  Indeed, only a very small value of signal-to-noise (S/N)$^2 = 0.3$ may be expected from the ongoing SPTPol experiment, although a much better signal-to-noise (S/N)$^2 = 28$ may be expected from the EPIC experiment.

\citet{su_yadav} also studied the possibility of reconstructing patchy reionization from CMB observables, and found that lensing induced non-gaussian features would produce a spurious $\tau(\hat n)$ signal much larger than the patchy $\tau(\hat n)$ expected from realistic reionization scenarios. It is however possible to simultaneously reconstruct the $\tau(\hat n)$ and the lensing potential $\Phi(\hat n)$ such that the $\tau(\hat n)$ estimate is not biased by lensing \citep{su_yadav}. Future experiments such as CMBPol can then detect patchy reionization with a signal-to-noise (S/N)$^2 \sim 100$. We note that a non-zero $EB$ correlation does not imply patchy reionization, as other physical effects such as primordial magnetic fields, can also induce a non-zero $EB$ correlation \citep{mag1, mag2}. 
 
In \citet{paper1} (henceforth Paper I), we described a new method for modeling inhomogeneous reionization on large scales. In the present article, we study the effect of inhomogeneous reionization (henceforth patchy $\tau$) on the CMB temperature power spectrum.  We present a technique that can detect patchy $\tau$ for root mean square $\rt$ as small as $\lesssim 0.003$, as seen in our maps. Our technique uses only temperature information which can be measured to very high precision by ongoing experiments such as ACT, SPT, ACTPol, and SPTPol. We consider a four point estimator in the coincidence limit rather than a quadratic estimator. Our patchy $\tau$ correlator relies on the fact that $\tau(\hat n)$ fluctuations are on scales far smaller than that of the primary CMB. The cross correlation between large scales and small scales of the CMB is therefore non-zero owing to the transfer of power due to patchy screening of the CMB. The only other physical effect (to our knowledge) that can induce a coupling across scales in this manner is CMB lensing. We account for CMB lensing and kSZ, and show that patchy $\tau$ can still be detected at high significance.  We then include  frequency dependent secondaries namely tSZ, CIB, and the radio background, and compute the patchy $\tau$ correlator. In the presence of these larger contaminants, patchy $\tau$ may be detected if the frequency dependent secondaries are minimized using observations at many different frequencies. We introduce gaussian noise to mimic residuals after frequency dependent secondaries are minimized, and show that a statistically significant detection of patchy $\tau$ requires that the residual noise level be $\lesssim$ 5 $\mu$K. We provide scaling relations for $\tau_{\rm RMS}$ as a function of the mean reionization redshift $\bar z$ and the duration of reionization $\Delta z$. Finally, we present our conclusions. We adopt the following cosmological parameters consistent with the WMAP observations \citep{wmap_data}: $\Omega_\Lambda = 0.73, \Omega_{\rm b} = 0.045, h = 0.7, n_{\rm s} = 0.96$, and $\sigma_8 = 0.80$.

\section{Secondary anisotropies due to patchy $\tau$} 

The CMB temperature as seen by an observer on earth in the direction $\hat n$  may be written as the sum of 2 terms:
\beq
T(\hat n) = \mathcal{T}_1(\hat n) + \mathcal{T}_2 (\hat n).
\eeq
$T(\hat n)$ represents the CMB brightness temperature in the Rayleigh-Jeans limit. $\mathcal{T}_1(\hat n)$ describes CMB photons that do not interact with free electrons, and therefore represent the temperature on the decoupling surface in the direction $\hat n$. Let $\tau(0, \hat n)$ be the optical depth due to scattering in the direction $\hat n$:
\beq
\tau(l, \hat n) = \sigma_{\rm T} \, \int_{l}^{l_\ast} dl' \, n_{\rm e} (l', \hat n),
\label{tau}
\eeq
where $dl = c \, dt$ is the proper distance along the line of sight. $l=0$ represents the observer, and $l_\ast$ is the decoupling surface. $n_{\rm e}(l, \hat n)$ is the free electron number density at $l$ in the direction $\hat n$, and $\sigma_{\rm T}$ is the Thomson scattering cross section. Let $\theta(\hat n)$ be the fractional change in the CMB temperature, and let $T_0$ be the CMB temperature averaged over all angles. $\mathcal{T}_1(\hat n)$ is then given by:
\beq
\mathcal{T}_1(\hat n) = T_0 \left [1 + \theta(\hat n) \right ] e^{-\tau(\hat n)} .
\eeq
The term $\mathcal{T}_2 (\hat n)$ describes CMB photons that scatter with free electrons. These photons  originate at different points on the decoupling surface,  and scatter into the line of sight $\hat n$. $\mathcal{T}_2 (\hat n)$ thus samples the entire decoupling surface, but depends on the line of sight due to the anisotropy in the ionized fraction.
\beq
\mathcal{T}_2(\hat n) = 2 T_0  \int_{0}^{l_\ast} dl \; n_{\rm e}(l, \hat n)  \;  \left \langle \frac{d\sigma}{d\mu} \; \left [ 1 + \theta(\hat n') \right ] e^{ {-\tau \left( \hat n', l \right ) }   } \right \rangle.
\eeq
The angle brackets indicate an average over different lines of sight. $\mu = \hat n \cdot \hat n'$, and the factor of 2 arises from the relation $ \int_{-1}^{1} d\mu \, f(\mu) = 2 \langle f(\mu) \rangle$. The differential cross section at low energies is:
\beq
d\sigma/d\mu = (3/8) \sigma_{\rm T} (1+\mu^2).
\eeq

Let us express the optical depth $\tau(\hat n)$ as the sum of an isotropic term $\tau_0 (l)$ and an angle dependent term $\eta(l, \hat n)$, i.e. $\tau(l, \hat n) = \tau_0(l) + \eta(l, \hat n)$. $\mathcal{T}_2 (\hat n)$ is then given by

\bea
\mathcal{T}_2 &\approx& \frac{3}{4} T_0 \, \sigma_{\rm T}   \int dl \; n_{\rm e}(l) e^{ -\tau_0(l) } \left \langle \left [ 1 + \mu^2 \right ] e^{-\eta \left( \hat n', l, l_\ast \right ) } \right \rangle \n
&=& T_0 \, \sigma_{\rm T} \int dl \, n_{\rm e}(l) e^{-\tau_0(l) } \; \left [ 1 + f_{1} + f_{2} \right ].
\label{ieq2}
\eea
In Equation (\ref{ieq2}), we have assumed that $\theta (\hat n)$ and $\eta (\hat n)$ are uncorrelated, and hence $\langle \theta \eta \rangle$ = 0. We have also ignored $\mu^2 \theta$ compared to $\mu^2$. The terms $f_1, f_2 \ll 1$ are given by
\begin{figure*}[!t]
\begin{center}
\scalebox{2.30}{\includegraphics{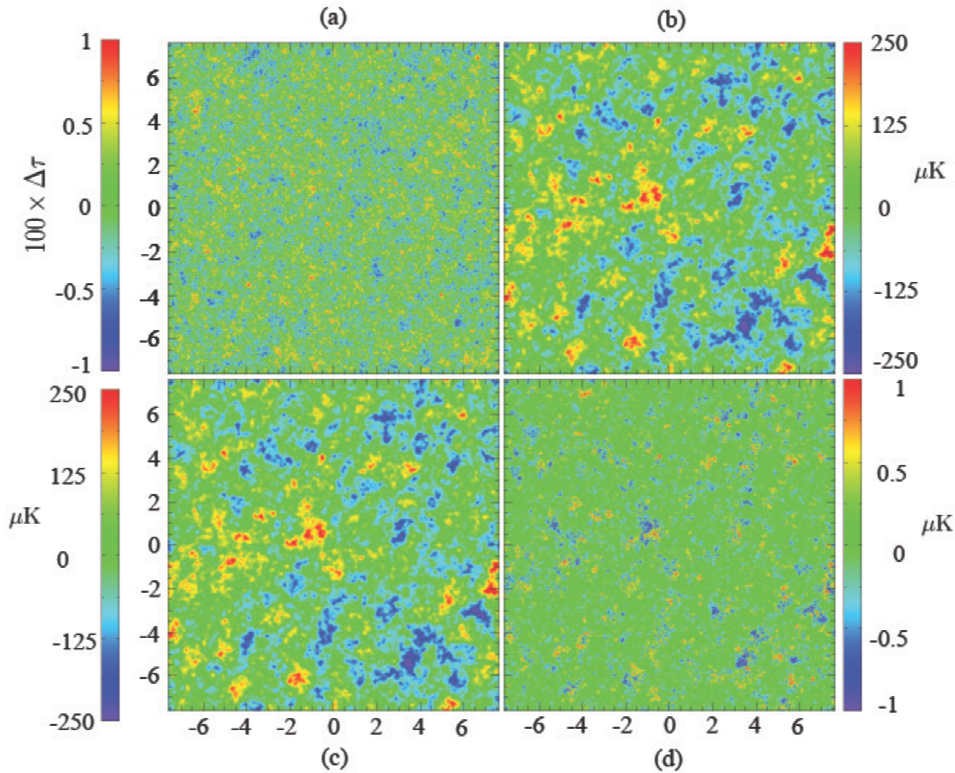}}
\end{center}
\caption{ Panel (a) shows fluctuations in the optical depth, in a $15^\circ \times 15^\circ$ region of the sky, from our maps. Panel (b) is a HEALPix realization of the primary CMB fluctuations. Panel (c) shows the CMB fluctuations with the effect of patchy $\tau$ included, while panel (d) shows only secondary CMB fluctuations due to the patchy $\tau$ effect. Fluctuations in the $\tau$ map are on much smaller scales than the CMB fluctuations. As seen from panels (b) and (c), the contribution due to patchy $\tau$ is small.
\label{fig1} }
\end{figure*}

\bea
f_1 &=& \frac{3}{4} \left [ \langle e^{-\eta} \rangle - 1 \right ] \approx \frac{3}{8} \langle \eta^2 \rangle + \cdots \n
f_2  &=& \frac{3}{4} \left [ \langle \mu^2 e^{-\eta} \rangle - \frac{1}{3} \right ] \approx -\frac{3}{4} \langle \mu^2 \eta \rangle + \cdots 
\eea

Typically, we find $\langle\eta^2 \rangle^{1/2}, \langle \mu^2 \eta \rangle \ll \tau_0$. From Equation (\ref{ieq2}) and Equation (\ref{tau}), we find to lowest order:

The CMB temperature in the direction $\hat n$ is
\beq
T(\hat n) = \mathcal{T}_1 + \mathcal{T}_2 = T_0 \left [ 1 + \theta(\hat n) \, e^{ -\tau(\hat n) } \right ].
\eeq

The fractional CMB temperature after scattering $\theta_{\rm obs} (\hat n)$ to lowest order is given by
\beq
\theta_{\rm obs} (\hat n) = \theta(\hat n)  \, e^ { -\tau(\hat n) }.
\label{theta_eqn}
\eeq

\beq
\mathcal{T}_2 = T_0 \left [ 1 - e^{ -\tau(\hat n) }  \right ].
\eeq

\noindent The total CMB $TT$ power spectrum may be obtained by performing a multipole decomposition of $\theta_{\rm obs} (\hat n)$. 

\begin{figure}[!b]
\begin{center}
\scalebox{0.50}{\includegraphics{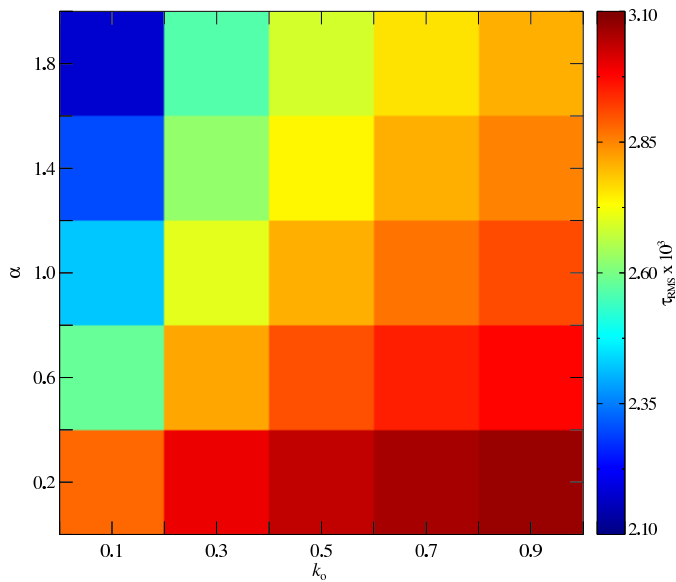}}
\end{center}
\caption{ $\rt \times 10^3$, shown for different values of $\alpha$ and $k_0$.  The redshift of 50\% reionization $\bar z$ = 10. $\rt < 4\%$ of the mean optical depth $\mt$ for $\bar z = 10$.
\label{fig2} }
\end{figure}

\begin{figure*}[!t]
\begin{center}
\scalebox{0.85}{\includegraphics [trim=0 50 0 350]  {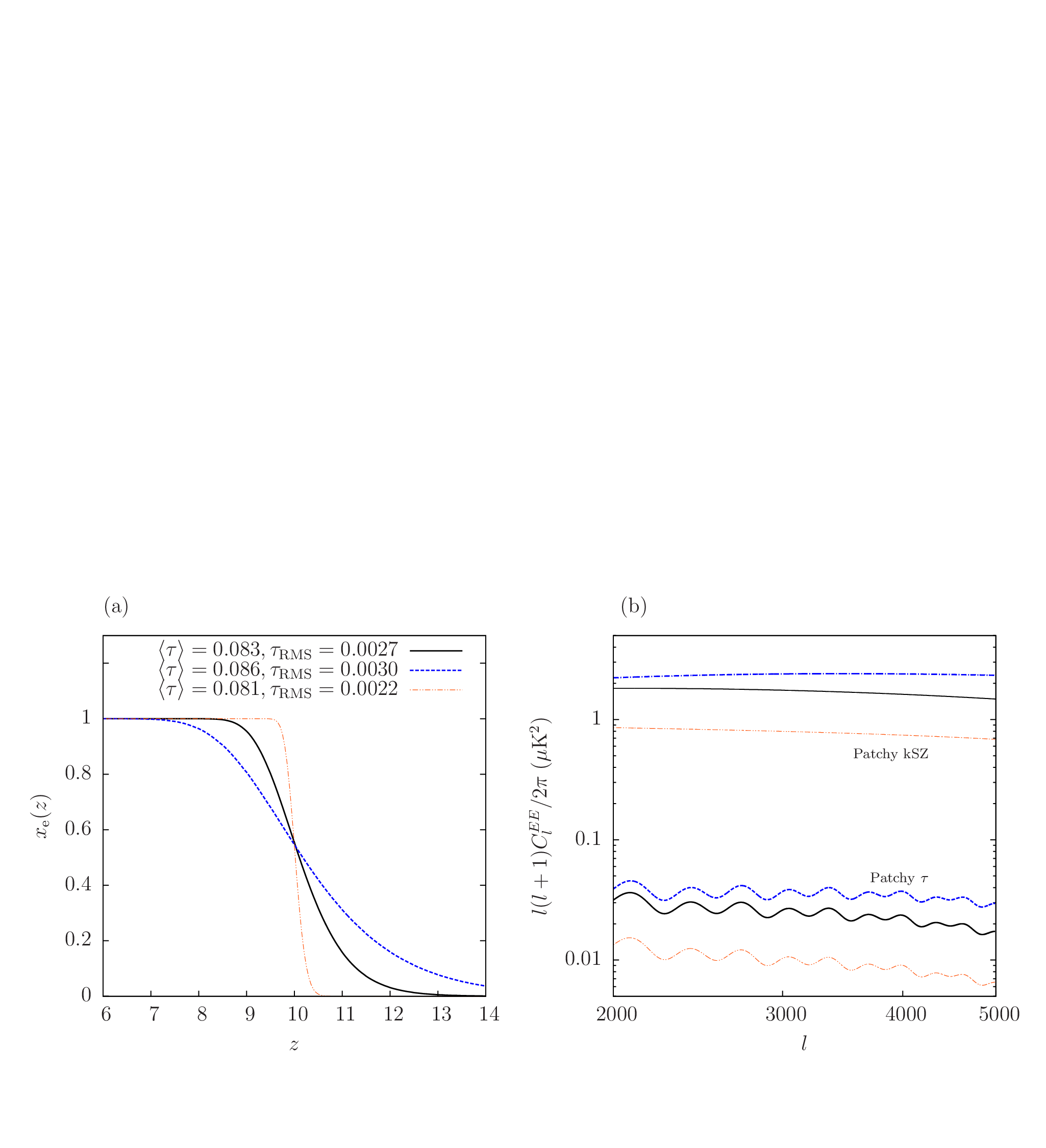}}
\end{center}
\caption{ Three different reionization models are shown (panel a) with the corresponding patchy $\tau$ and patchy kSZ ($z>5.5$) contributions (panel b). The homogeneous kSZ ($z<5.5$) is not shown. The patchy kSZ and patchy $\tau$ amplitudes will increase as $\langle \tau \rangle$ and $\tau_{\rm RMS}$ increase. However, the patchy kSZ is always significantly larger than the patchy $\tau$ contribution.
 \label{fig3} }
\end{figure*}

\section{Methodology}

We have developed a unique technique (described in Paper I) where the non-linear density field can be filtered with a simple parametric relation to directly obtain the reionization-redshift field. A particle-particle-particle-mesh (P$^3$M) N-body code is used to evolve $2048^3$ dark matter particles in a 2000 Mpc$/h$ box and generate nonlinear density fields $\rho(\vec{x})$ and velocity fields $\vec{v}(\vec{x})$ down to $z=5.5$.  Our semi-analytic model for reionization is based on results from radiative transfer hydrodynamic simulations \citep{trac_cen_loeb2008, paper1}. We track when a gas cell first becomes more than 90\% ionized and construct a 3D reionization-redshift field $z_{\rm RE}(\vec x)$ in parallel with the density field $\rho(\vec x)$.  Let us define the fluctuations in the matter density and the reionization-redshift fields as follows:

\bea
\delta_{\rm m}(\vec x) &=& \frac{\rho(\vec x) - \bar\rho }{\bar\rho} \n
\delta_{\rm z}(\vec x) &=& \frac{ \left[1 + z_{\rm RE}(\vec x) \right ] - \left[1 + \bar z \right ] }{1 + \bar z }.
\eea

From the Fourier transform of these quantities, we construct the bias and the cross-correlation functions:
\bea
b_{ \rm zm }(k) &=& \sqrt{ \frac{ \langle \delta_{\rm z} \delta_{\rm z} \rangle }{ \langle \delta_{\rm m} \delta_{\rm m} \rangle } }\n
r_{\rm zm} (k) &=& \frac{ \langle \delta_{\rm z} \delta_{\rm m} \rangle } { \sqrt{ \langle \delta_{\rm z} \delta_{\rm z} \rangle \langle \delta_{\rm m} \delta_{\rm m} \rangle } }
\eea

We find that the density and reionization fields are highly correlated on comoving scales $> 1$ Mpc/$h$ (see Paper I). $\delta_{\rm z}(k)$ may thus be obtained by applying a calibrated bias to $\delta_{\rm m}(k)$.

Full sky HEALPix (${\rm Nside}=4096$) maps of the patchy Thomson optical depth and kinetic Sunyaev-Zel'dovich effect are then constructed by raytracing through the simulated lightcone.  Panel (a) of Figure \ref{fig1} shows a $15^\circ  \times 15^\circ$ section of the $\tau$ map. The mean value of optical depth for this map is $\mt$ = 0.083, while the root mean square value $\rt$ = 0.0027. Panel (b) shows a realization of the CMB sky, computed using the HEALPix package\footnote{HEALPix may be downloaded from http://healpix.jpl.nasa.gov}\citep{healpix}. As can be seen, the $\tau$ fluctuations are on much smaller scales than the CMB fluctuations. Panel (c) shows the CMB fluctuations with patchy $\tau$ included, while panel (d) shows the contribution due to patchy $\tau$. From (b) and (c), we see that the patchy $\tau$ contribution is small.

 We use a simple parametric form for the bias function:
\begin{equation}
b_{ \rm zm} = \frac{\bo}{\left(1 + k/\kb\right)^{\al}}.
\label{bias}
\end{equation}
The bias approaches a constant on the largest scales $b_{\rm zm} \rightarrow 1/\delta_{\rm c} = 1/1.686$ \citep{barkana_loeb2004}. The bias parameter $b_{ {\rm zm} }$ contains 3 variables  $\bo$, $\kb$, and $\al$. $\bo$ is determined from analytical arguments in \citet{barkana_loeb2004}, while $\kb$, and $\al$ are found from simulations. Together with the redshift of 50\% reionization ($\bar z$), the parameters $\kb$ and $\al$ determine the degree of patchiness. Figure \ref{fig2} shows $\rt \times 10^3$ for different values of $\kb$ and $\al$, for a fixed value of $\bar z$ = 10. Thus, $\rt < 4\%$ of $\mt$, for $\bar z = 10$.

We consider three different reionization histories, shown in panel (a) of Figure \ref{fig3}. In all cases, the Universe is 50\% reionized at $\bar z$ = 10 (We choose $\bar z = 10$ for consistency with the WMAP measured value of optical depth).  The solid black curve shows our fiducial model, with $\mt$ = 0.083 and $\rt$ = 0.0027. The orange dot-dashed curve shows a shorter reionization history, with correspondingly smaller $\mt$ = 0.081 and $\rt$ = 0.0022. The dashed blue curve is plotted for an extended reionization model, with $\mt$ = 0.086 and $\rt$ = 0.0030. The panel on the right shows the secondary CMB power due to patchy $\tau$ compared with the secondary power due to patchy kSZ. The homogeneous kSZ power for $z < 5.5$ is the same for all reionization models, and is not shown here. We note that the patchy $\tau$ contribution is extremely small.

\begin{figure}[!t]
\begin{center}
\scalebox{0.95} {\includegraphics [trim=60 1 1 1] {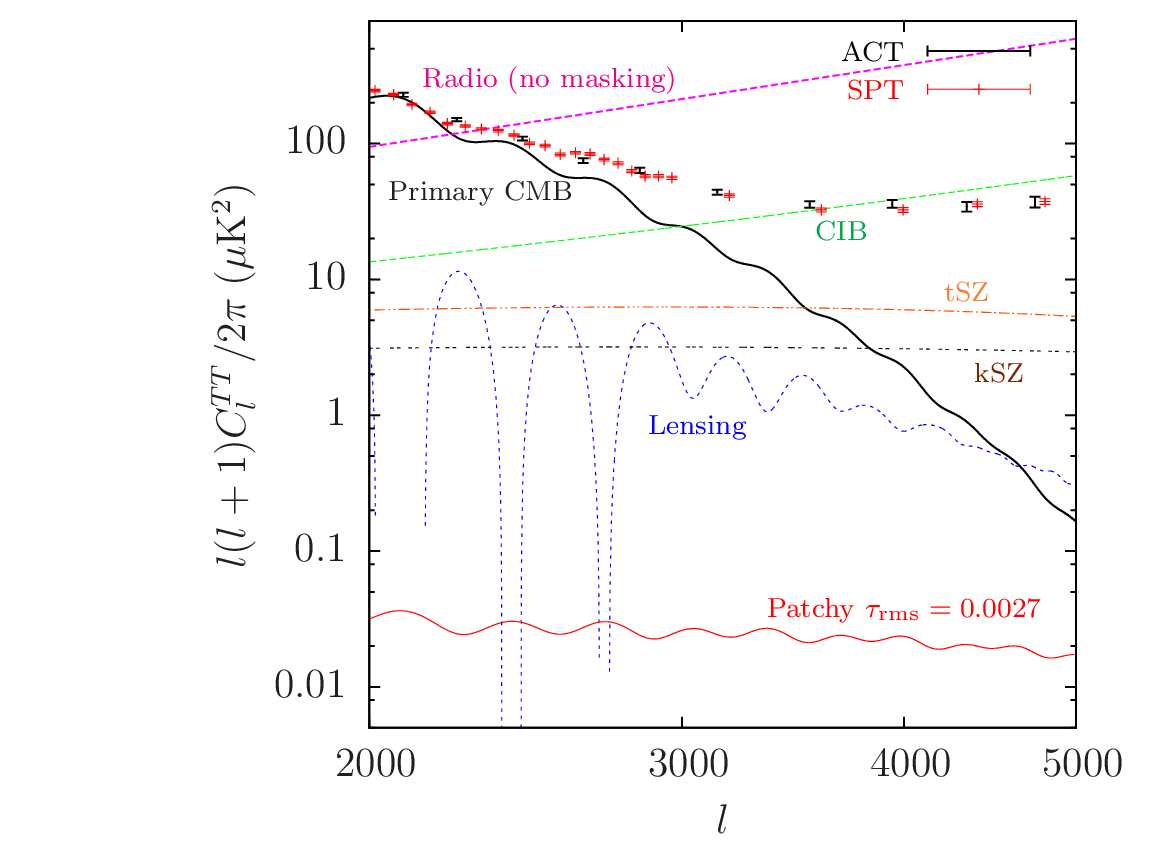}}
\end{center}
\caption{The primary CMB (solid, black) and  various secondary components (for our assumed cosmological parameters) at 148 GHz.  Also shown are the results from ACT (148 GHz) \citep{act_2013a} and SPT (150 GHz) \citep{spt2, story_etal}. The contribution from patchy $\tau$ is shown in red, for a reionization model with $\bar z =10, \rt = 0.0027$.
 \label{fig4} }
\end{figure}

Figure \ref{fig4} shows the different components of  the CMB temperature power spectrum, along with Atacama Cosmology Telescope (ACT) data \citep{act1, act_2013b, act3, act_2013a} at 148 GHz and South Pole Telescope (SPT) data \citep{spt2, keisler_etal_2011, story_etal} at 150 GHz. The solid black curve represents the primary CMB, while the solid red curve is the contribution due to patchy $\tau$, for a model characterized by a reionization redshift $\bar z$ = 10, mean value $\mt$ = 0.083, and patchiness $\rt = 0.0027$. Also shown are the contributions from patchy + homogeneous kSZ, the expected tSZ power (at 148 GHz), the contribution from CMB lensing, and the power from the CIB and the radio background (at 148 GHz). Figure \ref{fig4} shows the radio background before any masking is performed. The ACT collaboration has reported a residual radio power of $2.9 \pm 0.4 \, \mu$K$^2$ (defined at $l_0 = 3000, \nu_0 = 150$ GHz) after masking radio sources brighter than 15 mJy \citep{act_2013b}, while the SPT collaboration found a residual radio power of $1.28 \pm 0.19 \, \mu$K$^2$ ($l_0 = 3000, \nu_0 = 150$ GHz) with a masking threshold of 6.4 mJy \citep{spt2}. At multipole $l=3000$, the patchy kSZ ($z > 5.5$) contributes $\approx 1.7 \, \mu$K$^2$ (fiducial model) \citep{paper3}, the homogeneous kSZ ($z<5.5$) power is $\approx 1.4 \, \mu$K$^2$ \citep{trac_etal2011, bode_etal}, and the tSZ $\approx 6.3 \, \mu$K$^2$ \citep{trac_etal2011, bode_etal} (for our assumed cosmological parameters). In contrast, the patchy $\tau$ contribution is only $\approx 0.023 \, \mu$K$^2$.

In an attempt to constrain $\rt$, we perform a maximum likelihood analysis using CMB data from the ACT \citep{act1, act3, act_2013a}, SPT \citep{keisler_etal_2011, spt2, story_etal}, and WMAP \citep{wmap_data, wmap9a, wmap9b} experiments, and the publicly available CMB Boltzmann code  {\scriptsize CLASS} \citep{class1, class2}. We find that current CMB power spectrum data is insensitive to realistic values of patchiness $\tau_{\rm RMS} \lesssim 0.003$ because the patchy $\tau$ contribution is nearly 2 orders of magnitude smaller than the other secondary components such as the CIB, tSZ, and kSZ spectra. Moreover the uncertainty in the kSZ and tSZ templates exceeds the contribution due to patchy $\tau$. 

The ACTPol wide survey is expected to cover 4000 square degrees of the sky with a sensitivity of 20 $\mu$K-arcmin in temperature over a 1 year period, and a beam width $\approx$ 1.4 arcmin (148 GHz channel) \citep{actpol}. The ACTPol deep survey will cover five $2^\circ \times 15^\circ$ regions with a target sensitivity of 3 $\mu$K-arcminute \citep{actpol}. The ongoing SPTPol experiment is expected to cover $\sim$ 625 square degrees of the sky, with a target survey depth of 5 $\mu$K-arcmin over a 4 year period, and a beam width of $\sim$ 1 arcmin (150 GHz channel) \citep{sptpol1, sptpol2}. With these specifications, we expect the ACTPol and SPTPol experiments to be more sensitive to patchy kSZ, although it will still be very challenging to measure patchy $\tau$ using the CMB power spectra. We do not expect a significant improvement from the Planck mission, owing to the large ($\sim$ 5 - 7') beam width \citep{planck_hfi}, which limits CMB measurements to scales $l < 2000$. In the next section, we will discuss a much improved technique for detecting patchy reionization.

\section{Determining patchy $\tau$ from the temperature maps}

 We now discuss a different approach to detecting the presence of patchy reionization. We construct a simple estimator that is far more sensitive to non-zero $\rt$ than the power spectrum. Since the damping term $\exp -\tau(\hat n)$ multiplies the primary CMB temperature $T$ [Equation (\ref{theta_eqn})], the effect of patchy reionization is largest when $|T|$ is large. From Figures \ref{fig1} and \ref{fig4}, we see that patchy reionization alters the primary CMB only on very small scales. We therefore filter the CMB map (which includes the effect of patchy $\tau$) into 2 maps: (i) A map $f$ with information only on large scales, i.e. multipoles $l < l_{\rm boundary1}$, and (ii) A map $g$ with information only on small scales, $l > l_{\rm boundary2}$. Due to computational limitations, and to limit contamination of the small scale map by the large secondary anisotropies, we set $l_{\rm max} = 5000$ as the maximum multipole moment of interest. The 2 maps are then squared:
\bea
f &=&  T^2 (l < l_{\rm boundary1})  \n
g &=&  T^2 (l > l_{\rm boundary2}).
\label{eqn_fg}
\eea
 We compute the cross correlation $\corr$, which we call the patchy $\tau$ correlator.  $\delta f = f - \langle f \rangle$ and $\delta g = g - \langle g \rangle$ are fluctuations in the squared CMB temperature obtained from the filtered maps. The angle brackets denote an average over the map.  
 
Let us first examine a simple model wherein we ignore all secondaries besides patchy reionization. Let $\theta_{\rm obs}(\hat n)$ be the observed CMB temperature fluctuation, and let $\theta_{\rm cmb}(\hat n)$ be the primordial fluctuation (i.e. in the absence of Thomson scattering with free electrons). $\theta_{\rm cmb}$ consists of large scale and small scale modes, i.e. $\theta_{\rm cmb} = \theta_{\rm L} + \theta_{\rm S}$. 

The optical depth $\tau(\hat n)$ may be expressed as the sum of a constant term and a fluctuating term, i.e. $\tau(\hat n) = \langle\tau\rangle + \delta\tau(\hat n)$. The observed fluctuation $\theta_{\rm obs}(\hat n) = \theta_{\rm cmb} (\hat n) \times \exp -\delta\tau(\hat n) \approx \theta_{\rm cmb}(\hat n) - \delta\tau(\hat n) \theta_{\rm cmb}(\hat n)$, where we have dropped the constant term $\exp- \langle\tau\rangle$ because it is an overall multiplicative constant. The $\delta\tau(\hat n)$ fluctuations are on scales much smaller than the primary CMB fluctuations.  When the observed CMB map is filtered, we obtain a large scale map $\theta_{\rm L}$ and a small scale map $\theta_{\rm S} + \theta_{\rm L} \delta\tau$. The terms $\delta f$ and $\delta g$ are given by:
\bea
\delta f &=& \theta^2_{\rm L} - \langle \theta^2_{\rm L} \rangle \n
\delta g &=& \theta^2_{\rm S} - \langle \theta^2_{\rm S}  \rangle + \theta^2_{\rm L} \delta\tau^2 - \langle \theta^2_{\rm L} \delta\tau^2 \rangle + 2 \theta_{\rm L} \theta_{\rm S} \delta\tau.
\eea
The large scale and small scale modes of the CMB are independent of each other, and therefore uncorrelated. The $\delta\tau(\hat n)$ field is also uncorrelated with the CMB fluctuations. The patchy $\tau$ correlator is then computed as:
\beq
\langle \delta f \delta g \rangle = \langle \theta^4_{\rm L} \delta\tau^2 \rangle - \langle \theta^2_{\rm L} \rangle \langle \theta^2_{\rm L} \delta\tau^2 \rangle,
\eeq
with all other terms being zero. The patchy $\tau$ correlator is thus sensitive to optical depth fluctuations and vanishes in the limit of homogeneous reionization.
 
Let us now consider secondary anisotropies in addition to the primary CMB. We use the LensPix\footnote{LensPix available at http://cosmologist.info/lenspix} \citep{lenspix} (see also \citet{das_bode_2008} for a detailed description of CMB lensing) and HEALPix packages to generate 5 different realizations of the lensed and unlensed CMB. We  account for patchy reionization by multiplying the unlensed CMB by the damping term $\exp -\tau(\hat n)$ for a given reionization scenario. Secondary anisotropies such as CMB lensing, kSZ, tSZ, radio background, and the CIB are then added to the map. The combined map is filtered using HEALPix programs, to obtain a map with information only on large scales $l < l_{\rm boundary1}$, and a second map containing information only on small scales $l > l_{\rm boundary2}$. 

 Frequency dependent secondary effects may be reduced by observing the CMB at multiple frequencies. The Planck Satellite, expected to release data in early 2013 consists of a Low Frequency Instrument (LFI) \citep{planck_lfi} which observes the CMB at frequencies 30 GHz, 44 GHz, and 70 GHz, as well as a High Frequency Instrument (HFI) \citep{planck_hfi} sensitive to frequencies 100 GHz, 143 GHz, 217 GHz, 353 GHz, 545 GHz, and 857 GHz. The beam width varies from 33 arc minutes at 30 GHz to 5 arc minutes at 857 GHz \citep{planck_lfi, planck_hfi}. On relatively large scales, one could use observations at these 9 frequencies to remove frequency dependent secondary contributions. On small scales ($l > 3000$), one must use observations from other experiments with better angular resolution such as ACT, SPT, ACTPol, and SPTPol experiments. The ACT experiment is sensitive to frequencies 148 GHz and 219 GHz \citep{act1, act2, act_2013a}, while the SPT experiment measures the CMB at frequencies 95 GHz, 150 GHz, and 220 GHz \citep{keisler_etal_2011, spt2, story_etal}. 
 
Let us expand $\delta f$ and $\delta g$ into frequency independent and frequency dependent terms as:
\bea
\delta f &=& \delta f_0 = [f - \langle f \rangle]_0 \n
\delta g &=& \left (\delta g_0 + \delta g_\nu \right ) = [g - \langle g \rangle]_0 + [g - \langle g \rangle]_{\nu},
\label{f_and_g}
\eea
where the subscript 0 denotes a frequency independent contribution (the primary CMB, patchy $\tau$, lensing, and kSZ), while the subscript $\nu$ denotes a frequency dependent contribution (such as the tSZ effect, radio sources, and the CIB). In Equation (\ref{f_and_g}), we have made the assumption that frequency dependent terms can be removed from the large scale map using the many frequency channels of the Planck experiment. A similar assumption cannot be made for the small scale map however, as current and upcoming arcminute scale experiments measure the CMB at only 3 frequencies. We thus include the tSZ effect as well as the CIB and radio background in the small scale map. The patchy $\tau$ correlator $\corr$ may then be decomposed into frequency independent and frequency dependent terms:
\bea
\corr &=& \langle \delta f_0 \, \delta g_0 \rangle + \langle \delta f_0 \, \delta g_\nu \rangle  \n
&=& A + B(\nu).
\eea
The frequency independent term $A \ne 0$ since the patchy $\tau$ and CMB lensing contributions lead to a non-zero cross correlation between the squared large scale and small scale maps. The frequency dependent term $B(\nu)$ is zero since none of the frequency dependent secondaries are expected to be correlated with the primary CMB. In practice $B(\nu)$ will not be zero due to noise. By computing the cross correlation at different frequencies, it may be possible to overcome noise, thus minimizing frequency dependent contaminations.

\begin{figure*}[!t]
\begin{center}
\scalebox{1.35}{\includegraphics  [trim=100 330 150 10]  {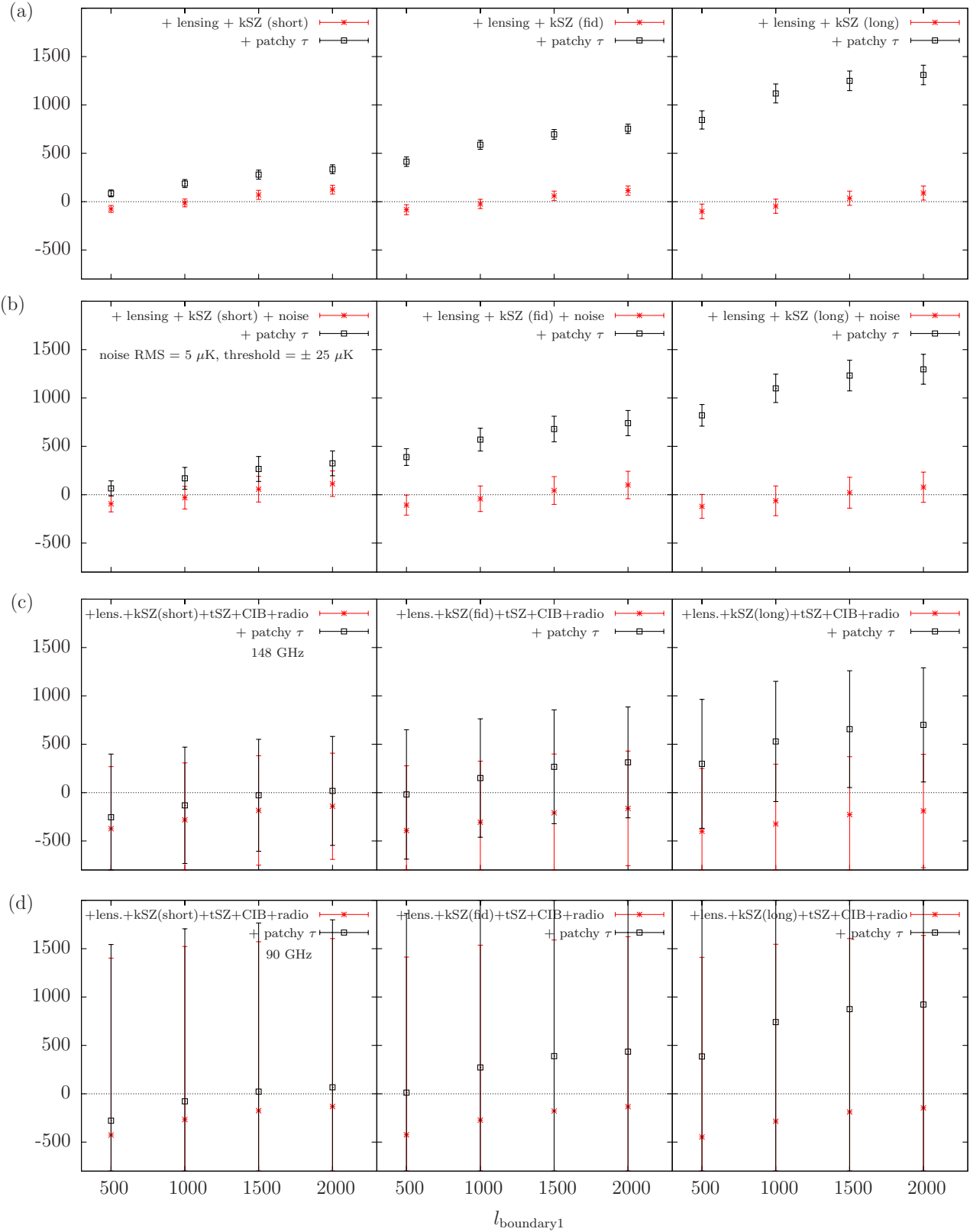}}
\end{center}
\caption{  The patchy $\tau$ correlator as a function of $l_{\rm boundary1}$. The top row (a) includes only frequency independent components, namely CMB lensing and kSZ, for the three different reionization histories considered in Figure \ref{fig3}. Patchy $\tau$ can be detected at high confidence, and the magnitude of the correlator can be used to constrain extended reionization histories. Row (b) shows the patchy $\tau$ correlator with gaussian noise included, to account for residuals after frequency dependent secondaries have been subtracted. Patchy $\tau$ may be detected if the residuals are smaller than $\approx$ 5 $\mu$K. Rows (c) and (d) include the tSZ, CIB, and radio background at frequencies 148 GHz and 90 GHz respectively, for the three different reionization histories.  A statistically significant detection of patchy $\tau$ can only be made if these large secondaries are minimized by a multi-frequency analysis. Zero correlation is shown for reference (thin broken line).
 \label{fig5} }
\end{figure*}

Figure \ref{fig5} shows the patchy $\tau$ correlator (in units of $\mu$K$^4$) with and without the effect of patchy $\tau$. $\corr$ is averaged over 5 CMB realizations, and plotted as a function of $l_{\rm boundary1}$ (largest multipole value for the large scale map). We set $l_{\rm boundary2} = 3000$ (smallest multipole value for the small scale map) since the patchy $\tau$ contribution is much smaller than the primary CMB at smaller multipoles, and the contribution from the CIB is very large at much larger multipoles. We choose $l_{\rm boundary1} \ne l_{\rm boundary2}$ in order to minimize contamination of the large scale map by secondary components. The error bars denote the root mean square (RMS) value of the different CMB realizations. 

The top row (a) of Figure \ref{fig5} shows the ideal scenario in which there exist only frequency independent contributions namely CMB lensing and kSZ. The three panels of row (a) show the three reionization histories considered in Figure \ref{fig3}: the short reionization scenario ($\langle\tau\rangle = 0.081, \tau_{\rm RMS} = 0.0022$), the fiducial model ($\langle\tau\rangle = 0.083, \tau_{\rm RMS} = 0.0027$), and the extended reionization history ($\langle\tau\rangle = 0.086, \tau_{\rm RMS} = 0.0030$). Since different multipoles of the primary CMB provide independent information, the correlation between large and small scale maps is zero for the primary CMB. Including the effect of CMB lensing however results in a non-zero cross correlation. This is expected as lensing of the CMB results in a redistribution of power, transferring CMB power from large scales to small scales. The cross correlation due to lensing is negative for small $l_{\rm boundary1}$, and increases, passing through zero at $l_{\rm boundary1} \sim 1200$. The patchy $\tau$ term is similarly correlated. The cross correlation due to patchy $\tau$ is always positive and increases with $l_{\rm boundary1}$. Moreover the patchy $\tau$ terms contributes significantly more to the cross correlation than the lensing term. Thus, with only the lensing and kSZ contributions, one can detect patchy reionization at high significance by computing the cross correlation between the squared large and small scale maps. However, we note that our simulations assume complete sky coverage ($f_{\rm sky} = 1$). In reality, even all-sky experiments such as Planck observe only $f_{\rm sky} \sim$ 0.65. Ground based experiments such as ACT and SPT obtain significantly smaller values of $f_{\rm sky}$, resulting in larger error bars $\propto f^{-1/2}_{\rm sky}$.

From the magnitude of the patchy $\tau$ correlator, we obtain information regarding the reionization history. It may thus be possible to distinguish different reionization models \emph {that predict the same mean optical depth} by computing the patchy $\tau$ correlator, provided the models predict different values of $\tau_{\rm RMS}$. This method is complementary to measuring the CMB polarization power spectrum which is sensitive to variations in $\langle \tau \rangle$, but cannot distinguish reionization models with different $\tau_{\rm RMS}$. We see no statistically significant cross correlation with the kSZ which merely serves to increase the size of the fluctuations about the mean.

\begin{figure*}[!t]
\begin{center}
\scalebox{0.94}{\includegraphics  [trim=35 520 0 80]  {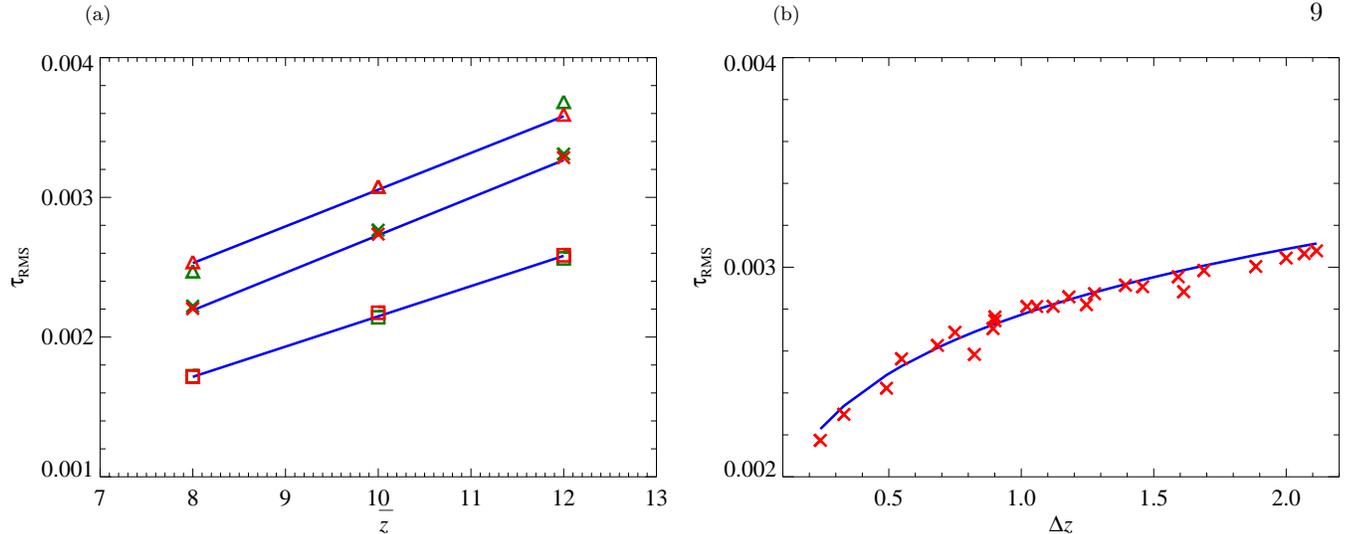}}
\end{center}
\caption{ The RMS fluctuation of the $\tau(\hat n)$ field. (a) shows $\tau_{\rm RMS}$ for different values of the mean reionization redshift $\bar z$, for values of $\Delta z$ = 0.2, 1.05, and 2.10 from bottom to top. The red symbols are the values calculated from the maps, the blue lines are the best fit, and green symbols are the values from the combined scaling law. (b) shows $\tau_{\rm RMS}$ as a function of the duration of reionization $\Delta z$, for a fixed $\bar z$ = 10. $\tau_{\rm RMS}$ is found to lie between 0.001 and 0.004 for most reionization histories. 
\label{fig6} }
\end{figure*}

The second row (b) shows the effect of adding gaussian distributed random noise with zero mean, and an RMS temperature of 5 $\mu$K/pixel to the small scale CMB maps. The purpose of adding noise is to mimic the effect of frequency dependent secondaries that are still present after a multi-frequency analysis has been employed to minimize the contribution from these components. This noise value is representative of the residuals from the tSZ and CIB contaminations, after bright pixels are masked out. A more thorough multi-frequency analysis of the tSZ and CIB contributions will likely yield smaller residuals.  Each pixel in our maps has an area of 0.74 arcmin$^2$. We apply a mask to the small scale map by setting to zero, any pixel whose temperature is greater than $+25 \, \mu$K or smaller than $-25 \, \mu$K. The threshold of $\pm 25 \, \mu$K minimizes the contribution from the added gaussian noise and the kSZ. No noise is added to the large scale maps since upcoming experiments such as the Planck mission are expected to measure the CMB temperature to high accuracy on scales $l \lesssim 2000$ using many frequency channels (cosmic variance is accounted for, by considering many realizations of the CMB sky).  The addition of gaussian noise does not affect the mean value of $\corr$, but it significantly increases the variance.  It is still possible to distinguish patchy $\tau$ with 5 $\mu$K of added noise. We thus need to ensure that secondary components and other sources of noise contribute $\lesssim 5 \, \mu$K for a detection of patchy $\tau$ with high significance.

The bottom two rows ((c) and (d)) study the effect of frequency dependent secondary effects in the small scale maps. We have included the contribution from the tSZ \citep{trac_etal2011, bode_etal}, CIB, and the radio background \citep{sehgal_etal2010} at frequencies 148 GHz (row (c)) and 90 GHz (row (d)), for the three reionization histories.\footnote{CIB and radio maps may be downloaded from \\ http://lambda.gsfc.nasa.gov/toolbox/tb\_cmbsim\_ov.cfm} As before, we apply a mask to the small scale map by setting to zero, any pixel whose temperature is greater than $+25 \, \mu$K or smaller than $-25 \, \mu$K.  This threshold value was chosen to be representative of the sensitivity of ongoing and upcoming CMB experiments. The ACT mission has achieved a noise level between 25 and 40 $\mu$K-arcmin \citep{act2}, while the noise level achieved by the SPT mission $\sim 18 \, \mu$K-arcmin. The ACTPol mission \citep{actpol} expects to achieve a sensitivity of $20 \,\mu$K-arcmin for the wide survey, and $3 \,\mu$K-arcmin for the deep survey at 148 GHz with 1 year of observation, while the SPTPol mission expects to achieve a noise level of 5 $\mu$K-arcmin for the 150 GHz channel \citep{sptpol2}, with 4 years of observation. (The pixel area in our maps is 0.74 arcmin$^2$). The noise in the relevant band, i.e. $l_{\rm boundary2} < l < l_{\rm max}$ may be smaller than the figure quoted by ACT and SPT, for scale independent noise). The frequency independent contributions due to CMB lensing and the kSZ are included in both large scale and small scale maps. We do not mask any pixel in the large scale map. 

As can be seen from rows (c) and (d), it is difficult to make a statistically significant detection of patchy $\tau$ when the CIB and the radio background are included.  At 148 GHz, the tSZ contributes an RMS temperature $\sim 4.2 \, \mu$K, while the CIB contributes $\sim 15 \, \mu$K (RMS temperatures are computed at all scales, the RMS in the restricted band is smaller). The magnitude of the patchy $\tau$ correlator is significantly decreased since there are many pixels that exceed the threshold of $\pm 25 \mu$K due to large discrete source poisson noise.  Row (d) show the computation at 90 GHz where the CIB is small. At a frequency of 90 GHz, the tSZ contributes an RMS temperature $\sim 6.8 \, \mu$K, while the CIB contributes $\sim 5.7 \, \mu$K. Unfortunately the radio background is prohibitively large at 90 GHz, and precludes a detection of patchy $\tau$ even for the extended reionization scenario. If bright radio galaxies can be detected and masked (without simultaneously masking the primary CMB) by using maps at multiple frequencies, one may hope to reduce the radio and CIB contributions down to $\sim$ few $\mu$K. The CIB contribution dramatically increases at still higher frequencies. At 219 GHz, the CIB contributes $\sim 49 \, \mu$K, completely overwhelming the small signal due to patchy $\tau$. Nevertheless, the tSZ contribution is nearly zero at this frequency, and observations at this frequency may be used to detect and minimize the tSZ contribution at lower frequencies.

It is possible to learn more about reionization science by computing the patchy $\tau$ correlator. We define the duration of reionization as follows:
\beq
\Delta z = z(x_{\rm e}=25\%) - z(x_{\rm e}=75\%).
\eeq
Figure \ref{fig6}  shows that patchiness in the reionization field increases with the mean reionization redshift $\bar z$, as well as the duration of reionization $\Delta z$.  Figure \ref{fig6} (a) shows $\tau_{\rm RMS}$ for different mean reionization redshifts, for values of $\Delta z$ = 0.2, 1.05, and 2.10  respectively (from bottom to top). Figure \ref{fig6} (b) shows how $\tau_{\rm RMS}$ scales with $\Delta z$ for fixed $\bar z$ = 10. Values of $\tau_{\rm RMS} \sim 0.001 - 0.004$ are well motivated for most reionization histories. The scaling laws are constrained to be:

\begin{equation}
\tau_{\rm  RMS} = 0.00275  \left[1.08\left(\frac{1 + \bar z}{11}\right) - 0.083\right],
\label{eq:zbar_cl}
\end{equation}

\noindent for a fixed $\Delta z = 1.05$ and

\begin{equation}
\tau_{\rm  RMS} = 0.00275 \left(\frac{\Delta z}{1.05}\right)^{0.15},
\label{eq:delz_cl}
\end{equation}

\noindent for a fixed $\bar z = 10$. The combination of Equations \ref{eq:zbar_cl} and \ref{eq:delz_cl} gives us

\begin{equation}
\tau_{\rm RMS} = 0.00297 \left[\left(\frac{1 + \bar z}{11}\right) -
0.0769\right] \left(\frac{\Delta z}{1.05}\right)^{0.15}.
\label{eq:fit_fun}
\end{equation}

 A statistically significant detection of patchy $\tau$ is possible (at low frequencies $\sim$ 90 GHz) for extended reionization histories with $\tau_{\rm RMS} \gtrsim 0.003$ and $\Delta z \gtrsim 2$ (Figure \ref{fig5} and Figure \ref{fig6}). Thus the patchy $\tau$ correlator in combination with the small scale CMB temperature power spectrum and the large scale polarization power spectrum can constrain extended reionization models.

\section{Conclusions}
We investigated the effect of patchy reionization on the CMB temperature. We showed that the anisotropy in the optical depth $\tau(\hat n)$ introduces secondary anisotropies in the CMB temperature on small scales. We analyzed current data from the WMAP and ACT observations, but found that the effect of patchy $\tau$ is too small to be seen in the $TT$ power spectrum. We then showed that there exists a simple estimator constructed from the temperature maps, which is sensitive to small values of patchiness. The key idea is that the damping caused by inhomogeneous Thomson scattering \emph{multiplies} the CMB temperature. Multiplication in angular space is equivalent to convolution in harmonic space which transfers power from low multipoles to high multipoles. Since most of the patchy $\tau$ contribution is on very small scales, we filtered the CMB map into low frequency and high frequency maps using HEALPix routines. We then computed the cross correlation between the squared maps, which we call the patchy $\tau$ correlator. We showed that the patchy $\tau$ component is clearly correlated with the primary CMB, while observations of the primary CMB on different scales are independent of each other. This technique is sensitive to patchiness values as small as $\tau_{\rm RMS} \sim 0.003$ seen in our maps.

We then considered the more difficult question of identifying patchy reionization in the presence of other secondary components. One might hope to remove frequency dependent secondaries through a multi-frequency analysis of CMB maps. The Planck satellite measures the CMB in 9 frequency channels for $2 < l \lesssim 2000$. The PIXIE (Primordial Inflation Explorer) satellite \citep{pixie} is a full sky experiment planned for 2017 that can observe the CMB on large scales. The instrument has 400 frequency channels from 30 GHz to 6 THz, and will greatly improve our understanding of frequency dependent contaminations on large scales. The ongoing ACT experiment uses 2 frequency channels for $515 < l < 9750$ \citep{act1, act_2013a}, while the SPT experiment uses 3 frequency channels for $675 < l < 9400$ \citep{keisler_etal_2011, spt2, story_etal}.  

The frequency independent contributions such as CMB lensing and kSZ are harder to remove. CMB lensing shows a non-zero correlation, but it is smaller than the correlation due to patchy $\tau$ and has a different dependence on the scale $l_{\rm boundary1}$. The kSZ shows no correlation with the primary CMB. We showed that patchy $\tau$ is easily detectable even when lensing and kSZ are included, particularly for extended reionization histories with large $\rt$. It may also be possible to distinguish different reionization models based on the magnitude of the patchy $\tau$ correlator, even when they have the same value of $\mt$, provided the $\rt$ values are different. On the other hand, the large angle $EE$ polarization may distinguish models with different $\mt$, but is insensitive to $\rt$. The patchy $\tau$ correlator is thus a useful probe of reionization that provides information complementary to what may be obtained from the polarization power spectrum. With future data sets, one may hope to distinguish between different reionization histories by computing the cross correlation parameter in conjunction with the $EE$ and the kSZ power spectra (a detailed treatment of the polarization and patchy kSZ power spectra is left to a companion paper \citep{paper3}).

We then included the tSZ, the radio background, and the CIB to the maps, and attempted to identify the patchy $\tau$ signal. Detection of patchy $\tau$ is possible if these large contaminants are removed or minimized by means of a multi-frequency analysis. We included gaussian random noise in addition to lensing and kSZ in order to account for frequency dependent secondaries that are still present after a multi-frequency analysis has been employed to minimize the contribution from these components. We showed that a detection of patchy $\tau$ would require cleaning the small scale maps to a noise level $\sim$ 5 $\mu$K/pixel. Such a low noise level would require a good understanding of the instrument as well as observations of the CMB at multiple frequencies in order to minimize frequency dependent contaminations. We also showed that computing $\tau_{\rm RMS}$ can set bounds on the duration of reionization. Finally, we provided scaling relations for $\tau_{\rm RMS}$ as a function of the mean reionization redshift $\bar z$ and the duration of reionization $\Delta z$.

 Current and upcoming experiments are sensitive to low frequency channels and should be able to constrain extended reionization histories. Quantifying the patchiness of reionization could provide information regarding ionizing sources. \citet{visbal_loeb2012} found that when ionization by X-rays is included, the patchy contribution is reduced owing to the larger mean free path of the ionizing photons. For a fixed mean optical depth $\mt$, this implies a smaller reionization contribution from lower energy sources, and hence smaller values of $\rt$, influencing the patchy kSZ \citep{visbal_loeb2012}, as well as the patchy $\tau$ correlator. Early reionization from other sources such as dark matter annihilation \citep{natarajan_schwarz1, natarajan_schwarz2, natarajan_schwarz3} may also be probed in a similar way \citep{visbal_loeb2012}. A more complete analysis is left to future work.

\acknowledgments{A.N. and N.B. are supported by a McWilliams postdoctoral fellowship awarded by the Bruce and Astrid McWilliams Center for Cosmology. We thank Brad Benson, Joanna Dunkley, Yu Feng, Michael Niemack, Christian Reichardt, Jonathan Sievers, Kendrick Smith, and David Spergel for helpful discussions. A.N. is supported in part by NSF grant AST-1211777.  H.T. is supported in part by NSF grant AST-1109730.  A.L. is supported in part by NSF grant AST-0907890, and NASA grants NNX08AL43G and NNA09DB30A.  The simulations were performed at the Pittsburgh Supercomputing Center (PSC) and the Princeton Institute for Computational Science and Engineering (PICSciE). We thank Roberto Gomez and Rick Costa at the PSC and Bill Wichser at PICSciE for invaluable help with computing. Some of the results in this paper have been derived using the HEALPix (K.M. G\'{o}rski et al. 2005) and LensPix (A. Lewis 2005) packages. We acknowledge the use of the Legacy Archive for Microwave Background Data Analysis (LAMBDA). Support for LAMBDA is provided by the NASA Office of Space Science. }

\bibliography{references}

\begin{thebibliography}{60}
\expandafter\ifx\csname natexlab\endcsname\relax\def\natexlab#1{#1}\fi

\bibitem[{{Austermann} {et~al.}(2012){Austermann}, {Aird}, {Beall}, {Becker},
  {Bender}, {Benson}, {Bleem}, {Britton}, {Carlstrom}, {Chang}, {Chiang},
  {Cho}, {Crawford}, {Crites}, {Datesman}, {de Haan}, {Dobbs}, {George},
  {Halverson}, {Harrington}, {Henning}, {Hilton}, {Holder}, {Holzapfel},
  {Hoover}, {Huang}, {Hubmayr}, {Irwin}, {Keisler}, {Kennedy}, {Knox}, {Lee},
  {Leitch}, {Li}, {Lueker}, {Marrone}, {McMahon}, {Mehl}, {Meyer}, {Montroy},
  {Natoli}, {Nibarger}, {Niemack}, {Novosad}, {Padin}, {Pryke}, {Reichardt},
  {Ruhl}, {Saliwanchik}, {Sayre}, {Schaffer}, {Shirokoff}, {Stark}, {Story},
  {Vanderlinde}, {Vieira}, {Wang}, {Williamson}, {Yefremenko}, {Yoon}, \&
  {Zahn}}]{sptpol2}
{Austermann}, J.~E. {et~al.} 2012, in Society of Photo-Optical Instrumentation
  Engineers (SPIE) Conference Series, Vol. 8452, Society of Photo-Optical
  Instrumentation Engineers (SPIE) Conference Series

\bibitem[{{Barkana} \& {Loeb}(2001)}]{barkana_loeb2001}
{Barkana}, R., \& {Loeb}, A. 2001, \physrep, 349, 125

\bibitem[{{Barkana} \& {Loeb}(2004)}]{barkana_loeb2004}
---. 2004, \apj, 609, 474

\bibitem[{{Battaglia} {et~al.}(2013){Battaglia}, {Natarajan}, {Trac}, {Cen}, \&
  {Loeb}}]{paper3}
{Battaglia}, N., {Natarajan}, A., {Trac}, H., {Cen}, R., \& {Loeb}, A. 2013,
  \apj, 776, 83

\bibitem[{Battaglia {et~al.}(2013)Battaglia, Trac, Cen, \& Loeb}]{paper1}
Battaglia, N., Trac, H., Cen, R., \& Loeb, A. 2013, \apj, 776, 81

\bibitem[{{Becker} {et~al.}(2001){Becker}, {Fan}, {White}, {Strauss},
  {Narayanan}, {Lupton}, {Gunn}, {Annis}, {Bahcall}, {Brinkmann}, {Connolly},
  {Csabai}, {Czarapata}, {Doi}, {Heckman}, {Hennessy}, {Ivezi{\'c}}, {Knapp},
  {Lamb}, {McKay}, {Munn}, {Nash}, {Nichol}, {Pier}, {Richards}, {Schneider},
  {Stoughton}, {Szalay}, {Thakar}, \& {York}}]{becker_etal2001}
{Becker}, R.~H. {et~al.} 2001, \aj, 122, 2850

\bibitem[{{Bennett} {et~al.}(2012){Bennett}, {Larson}, {Weiland}, {Jarosik},
  {Hinshaw}, {Odegard}, {Smith}, {Hill}, {Gold}, {Halpern}, {Komatsu}, {Nolta},
  {Page}, {Spergel}, {Wollack}, {Dunkley}, {Kogut}, {Limon}, {Meyer}, {Tucker},
  \& {Wright}}]{wmap9b}
{Bennett}, C.~L. {et~al.} 2012, arXiv:1212.5225

\bibitem[{{Blas} {et~al.}(2011){Blas}, {Lesgourgues}, \& {Tram}}]{class2}
{Blas}, D., {Lesgourgues}, J., \& {Tram}, T. 2011, \jcap, 7, 34

\bibitem[{{Bleem} {et~al.}(2012){Bleem}, {Ade}, {Aird}, {Austermann}, {Beall},
  {Becker}, {Benson}, {Britton}, {Carlstrom}, {Chang}, {Cho}, {de Haan},
  {Crawford}, {Crites}, {Datesman}, {Dobbs}, {Everett}, {Ewall-Wice}, {George},
  {Halverson}, {Harrington}, {Henning}, {Hilton}, {Holzapfel}, {Hoover},
  {Hubmayr}, {Irwin}, {Keisler}, {Kennedy}, {Lee}, {Leitch}, {Li}, {Lueker},
  {Marrone}, {McMahon}, {Mehl}, {Meyer}, {Montgomery}, {Montroy}, {Natoli},
  {Nibarger}, {Niemack}, {Novosad}, {Padin}, {Pryke}, {Reichardt}, {Ruhl},
  {Saliwanchik}, {Sayre}, {Schafer}, {Shirokoff}, {Story}, {Vanderlinde},
  {Vieira}, {Wang}, {Williamson}, {Yefremenko}, {Yoon}, \& {Young}}]{sptpol1}
{Bleem}, L. {et~al.} 2012, Journal of Low Temperature Physics, 167, 859

\bibitem[{{Bode} {et~al.}(2012){Bode}, {Ostriker}, {Cen}, \&
  {Trac}}]{bode_etal}
{Bode}, P., {Ostriker}, J.~P., {Cen}, R., \& {Trac}, H. 2012, arXiv:1204.1762

\bibitem[{{Ciardi} {et~al.}(2003){Ciardi}, {Ferrara}, \&
  {White}}]{ciardi_etal2003}
{Ciardi}, B., {Ferrara}, A., \& {White}, S.~D.~M. 2003, \mnras, 344, L7

\bibitem[{{Das} \& {Bode}(2008)}]{das_bode_2008}
{Das}, S., \& {Bode}, P. 2008, \apj, 682, 1

\bibitem[{{Das} {et~al.}(2013){Das}, {Louis}, {Nolta}, {Addison},
  {Battistelli}, {Bond}, {Calabrese}, {Devlin}, {Dicker}, {Dunkley},
  {D{\"u}nner}, {Fowler}, {Gralla}, {Hajian}, {Halpern}, {Hasselfield},
  {Hilton}, {Hincks}, {Hlozek}, {Huffenberger}, {Hughes}, {Irwin}, {Kosowsky},
  {Lupton}, {Marriage}, {Marsden}, {Menanteau}, {Moodley}, {Niemack}, {Page},
  {Partridge}, {Reese}, {Schmitt}, {Sehgal}, {Sherwin}, {Sievers}, {Spergel},
  {Staggs}, {Swetz}, {Switzer}, {Thornton}, {Trac}, \& {Wollack}}]{act_2013a}
{Das}, S. {et~al.} 2013, arXiv:1301.1037

\bibitem[{{Das} {et~al.}(2011){Das}, {Marriage}, {Ade}, {Aguirre}, {Amiri},
  {Appel}, {Barrientos}, {Battistelli}, {Bond}, {Brown}, {Burger}, {Chervenak},
  {Devlin}, {Dicker}, {Bertrand Doriese}, {Dunkley}, {D{\"u}nner},
  {Essinger-Hileman}, {Fisher}, {Fowler}, {Hajian}, {Halpern}, {Hasselfield},
  {Hern{\'a}ndez-Monteagudo}, {Hilton}, {Hilton}, {Hincks}, {Hlozek},
  {Huffenberger}, {Hughes}, {Hughes}, {Infante}, {Irwin}, {Baptiste Juin},
  {Kaul}, {Klein}, {Kosowsky}, {Lau}, {Limon}, {Lin}, {Lupton}, {Marsden},
  {Martocci}, {Mauskopf}, {Menanteau}, {Moodley}, {Moseley}, {Netterfield},
  {Niemack}, {Nolta}, {Page}, {Parker}, {Partridge}, {Reid}, {Sehgal},
  {Sherwin}, {Sievers}, {Spergel}, {Staggs}, {Swetz}, {Switzer}, {Thornton},
  {Trac}, {Tucker}, {Warne}, {Wollack}, \& {Zhao}}]{act3}
---. 2011, \apj, 729, 62

\bibitem[{{Dor{\'e}} {et~al.}(2007){Dor{\'e}}, {Holder}, {Alvarez}, {Iliev},
  {Mellema}, {Pen}, \& {Shapiro}}]{dore2007}
{Dor{\'e}}, O., {Holder}, G., {Alvarez}, M., {Iliev}, I.~T., {Mellema}, G.,
  {Pen}, U.-L., \& {Shapiro}, P.~R. 2007, \prd, 76, 043002

\bibitem[{{Dunkley} {et~al.}(2013){Dunkley}, {Calabrese}, {Sievers}, {Addison},
  {Battaglia}, {Battistelli}, {Bond}, {Das}, {Devlin}, {D{\"u}nner}, {Fowler},
  {Gralla}, {Hajian}, {Halpern}, {Hasselfield}, {Hincks}, {Hlozek}, {Hughes},
  {Irwin}, {Kosowsky}, {Louis}, {Marriage}, {Marsden}, {Menanteau}, {Moodley},
  {Niemack}, {Nolta}, {Page}, {Partridge}, {Sehgal}, {Spergel}, {Staggs},
  {Switzer}, {Trac}, \& {Wollack}}]{act_2013b}
{Dunkley}, J. {et~al.} 2013, \jcap, 7, 25

\bibitem[{{Dunkley} {et~al.}(2011){Dunkley}, {Hlozek}, {Sievers}, {Acquaviva},
  {Ade}, {Aguirre}, {Amiri}, {Appel}, {Barrientos}, {Battistelli}, {Bond},
  {Brown}, {Burger}, {Chervenak}, {Das}, {Devlin}, {Dicker}, {Bertrand
  Doriese}, {D{\"u}nner}, {Essinger-Hileman}, {Fisher}, {Fowler}, {Hajian},
  {Halpern}, {Hasselfield}, {Hern{\'a}ndez-Monteagudo}, {Hilton}, {Hilton},
  {Hincks}, {Huffenberger}, {Hughes}, {Hughes}, {Infante}, {Irwin}, {Juin},
  {Kaul}, {Klein}, {Kosowsky}, {Lau}, {Limon}, {Lin}, {Lupton}, {Marriage},
  {Marsden}, {Mauskopf}, {Menanteau}, {Moodley}, {Moseley}, {Netterfield},
  {Niemack}, {Nolta}, {Page}, {Parker}, {Partridge}, {Reid}, {Sehgal},
  {Sherwin}, {Spergel}, {Staggs}, {Swetz}, {Switzer}, {Thornton}, {Trac},
  {Tucker}, {Warne}, {Wollack}, \& {Zhao}}]{act1}
---. 2011, \apj, 739, 52

\bibitem[{{Dunkley} {et~al.}(2009){Dunkley}, {Komatsu}, {Nolta}, {Spergel},
  {Larson}, {Hinshaw}, {Page}, {Bennett}, {Gold}, {Jarosik}, {Weiland},
  {Halpern}, {Hill}, {Kogut}, {Limon}, {Meyer}, {Tucker}, {Wollack}, \&
  {Wright}}]{dunkley_etal2009}
---. 2009, \apjs, 180, 306

\bibitem[{{Dvorkin} {et~al.}(2009){Dvorkin}, {Hu}, \&
  {Smith}}]{dvorkin_etal2009}
{Dvorkin}, C., {Hu}, W., \& {Smith}, K.~M. 2009, \prd, 79, 107302

\bibitem[{{Dvorkin} \& {Smith}(2009)}]{dvorkin_smith2009}
{Dvorkin}, C., \& {Smith}, K.~M. 2009, \prd, 79, 043003

\bibitem[{{Fan} {et~al.}(2001){Fan}, {Narayanan}, {Lupton}, {Strauss}, {Knapp},
  {Becker}, {White}, {Pentericci}, {Leggett}, {Haiman}, {Gunn}, {Ivezi{\'c}},
  {Schneider}, {Anderson}, {Brinkmann}, {Bahcall}, {Connolly}, {Csabai}, {Doi},
  {Fukugita}, {Geballe}, {Grebel}, {Harbeck}, {Hennessy}, {Lamb}, {Miknaitis},
  {Munn}, {Nichol}, {Okamura}, {Pier}, {Prada}, {Richards}, {Szalay}, \&
  {York}}]{fan_etal2001}
{Fan}, X. {et~al.} 2001, \aj, 122, 2833

\bibitem[{{Fan} {et~al.}(2002){Fan}, {Narayanan}, {Strauss}, {White}, {Becker},
  {Pentericci}, \& {Rix}}]{fan_etal2002}
{Fan}, X., {Narayanan}, V.~K., {Strauss}, M.~A., {White}, R.~L., {Becker},
  R.~H., {Pentericci}, L., \& {Rix}, H.-W. 2002, \aj, 123, 1247

\bibitem[{{Gluscevic} {et~al.}(2013){Gluscevic}, {Kamionkowski}, \&
  {Hanson}}]{vera_etal}
{Gluscevic}, V., {Kamionkowski}, M., \& {Hanson}, D. 2013, \prd, 87, 047303

\bibitem[{{G{\'o}rski} {et~al.}(2005){G{\'o}rski}, {Hivon}, {Banday},
  {Wandelt}, {Hansen}, {Reinecke}, \& {Bartelmann}}]{healpix}
{G{\'o}rski}, K.~M., {Hivon}, E., {Banday}, A.~J., {Wandelt}, B.~D., {Hansen},
  F.~K., {Reinecke}, M., \& {Bartelmann}, M. 2005, \apj, 622, 759

\bibitem[{{Gunn} \& {Peterson}(1965)}]{gunn_peterson1965}
{Gunn}, J.~E., \& {Peterson}, B.~A. 1965, \apj, 142, 1633

\bibitem[{{Hinshaw} {et~al.}(2012){Hinshaw}, {Larson}, {Komatsu}, {Spergel},
  {Bennett}, {Dunkley}, {Nolta}, {Halpern}, {Hill}, {Odegard}, {Page}, {Smith},
  {Weiland}, {Gold}, {Jarosik}, {Kogut}, {Limon}, {Meyer}, {Tucker}, {Wollack},
  \& {Wright}}]{wmap9a}
{Hinshaw}, G. {et~al.} 2012, arXiv:1212.5226

\bibitem[{{Hlozek} {et~al.}(2012){Hlozek}, {Dunkley}, {Addison}, {Appel},
  {Bond}, {Sofia Carvalho}, {Das}, {Devlin}, {D{\"u}nner}, {Essinger-Hileman},
  {Fowler}, {Gallardo}, {Hajian}, {Halpern}, {Hasselfield}, {Hilton}, {Hincks},
  {Hughes}, {Irwin}, {Klein}, {Kosowsky}, {Marriage}, {Marsden}, {Menanteau},
  {Moodley}, {Niemack}, {Nolta}, {Page}, {Parker}, {Partridge}, {Rojas},
  {Sehgal}, {Sherwin}, {Sievers}, {Spergel}, {Staggs}, {Swetz}, {Switzer},
  {Thornton}, \& {Wollack}}]{act2}
{Hlozek}, R. {et~al.} 2012, \apj, 749, 90

\bibitem[{{Hu}(2000)}]{hu2000}
{Hu}, W. 2000, \apj, 529, 12

\bibitem[{{Kahniashvili} \& {Ratra}(2005)}]{mag1}
{Kahniashvili}, T., \& {Ratra}, B. 2005, \prd, 71, 103006

\bibitem[{{Keisler} {et~al.}(2011){Keisler}, {Reichardt}, {Aird}, {Benson},
  {Bleem}, {Carlstrom}, {Chang}, {Cho}, {Crawford}, {Crites}, {de Haan},
  {Dobbs}, {Dudley}, {George}, {Halverson}, {Holder}, {Holzapfel}, {Hoover},
  {Hou}, {Hrubes}, {Joy}, {Knox}, {Lee}, {Leitch}, {Lueker}, {Luong-Van},
  {McMahon}, {Mehl}, {Meyer}, {Millea}, {Mohr}, {Montroy}, {Natoli}, {Padin},
  {Plagge}, {Pryke}, {Ruhl}, {Schaffer}, {Shaw}, {Shirokoff}, {Spieler},
  {Staniszewski}, {Stark}, {Story}, {van Engelen}, {Vanderlinde}, {Vieira},
  {Williamson}, \& {Zahn}}]{keisler_etal_2011}
{Keisler}, R. {et~al.} 2011, \apj, 743, 28

\bibitem[{{Kogut} {et~al.}(2011){Kogut}, {Fixsen}, {Chuss}, {Dotson}, {Dwek},
  {Halpern}, {Hinshaw}, {Meyer}, {Moseley}, {Seiffert}, {Spergel}, \&
  {Wollack}}]{pixie}
{Kogut}, A. {et~al.} 2011, \jcap, 7, 25

\bibitem[{{Kristiansen} \& {Ferreira}(2008)}]{mag2}
{Kristiansen}, J.~R., \& {Ferreira}, P.~G. 2008, \prd, 77, 123004

\bibitem[{{Lamarre} {et~al.}(2010){Lamarre}, {Puget}, {Ade}, {Bouchet},
  {Guyot}, {Lange}, {Pajot}, {Arondel}, {Benabed}, {Beney}, {Beno{\^i}t},
  {Bernard}, {Bhatia}, {Blanc}, {Bock}, {Br{\'e}elle}, {Bradshaw}, {Camus},
  {Catalano}, {Charra}, {Charra}, {Church}, {Couchot}, {Coulais}, {Crill},
  {Crook}, {Dassas}, {de Bernardis}, {Delabrouille}, {de Marcillac}, {Delouis},
  {D{\'e}sert}, {Dumesnil}, {Dupac}, {Efstathiou}, {Eng}, {Evesque},
  {Fourmond}, {Ganga}, {Giard}, {Gispert}, {Guglielmi}, {Haissinski},
  {Henrot-Versill{\'e}}, {Hivon}, {Holmes}, {Jones}, {Koch}, {Lagard{\`e}re},
  {Lami}, {Land{\'e}}, {Leriche}, {Leroy}, {Longval},
  {Mac{\'{\i}}as-P{\'e}rez}, {Maciaszek}, {Maffei}, {Mansoux}, {Marty}, {Masi},
  {Mercier}, {Miville-Desch{\^e}nes}, {Moneti}, {Montier}, {Murphy},
  {Narbonne}, {Nexon}, {Paine}, {Pahn}, {Perdereau}, {Piacentini}, {Piat},
  {Plaszczynski}, {Pointecouteau}, {Pons}, {Ponthieu}, {Prunet}, {Rambaud},
  {Recouvreur}, {Renault}, {Ristorcelli}, {Rosset}, {Santos}, {Savini},
  {Serra}, {Stassi}, {Sudiwala}, {Sygnet}, {Tauber}, {Torre}, {Tristram},
  {Vibert}, {Woodcraft}, {Yurchenko}, \& {Yvon}}]{planck_hfi}
{Lamarre}, J.-M. {et~al.} 2010, \aap, 520, A9

\bibitem[{{Larson} {et~al.}(2011{\natexlab{a}}){Larson}, {Dunkley}, {Hinshaw},
  {Komatsu}, {Nolta}, {Bennett}, {Gold}, {Halpern}, {Hill}, {Jarosik}, {Kogut},
  {Limon}, {Meyer}, {Odegard}, {Page}, {Smith}, {Spergel}, {Tucker}, {Weiland},
  {Wollack}, \& {Wright}}]{larson_etal2011}
{Larson}, D. {et~al.} 2011{\natexlab{a}}, \apjs, 192, 16

\bibitem[{{Larson} {et~al.}(2011{\natexlab{b}}){Larson}, {Dunkley}, {Hinshaw},
  {Komatsu}, {Nolta}, {Bennett}, {Gold}, {Halpern}, {Hill}, {Jarosik}, {Kogut},
  {Limon}, {Meyer}, {Odegard}, {Page}, {Smith}, {Spergel}, {Tucker}, {Weiland},
  {Wollack}, \& {Wright}}]{wmap_data}
---. 2011{\natexlab{b}}, \apjs, 192, 16

\bibitem[{{Lesgourgues}(2011)}]{class1}
{Lesgourgues}, J. 2011, arXiv:1104.2932

\bibitem[{{Lewis}(2005)}]{lenspix}
{Lewis}, A. 2005, \prd, 71, 083008

\bibitem[{{Liu} {et~al.}(2001){Liu}, {Sugiyama}, {Benson}, {Lacey}, \&
  {Nusser}}]{liu_etal2001}
{Liu}, G.-C., {Sugiyama}, N., {Benson}, A.~J., {Lacey}, C.~G., \& {Nusser}, A.
  2001, \apj, 561, 504

\bibitem[{{Loeb} \& {Barkana}(2001)}]{loeb_barkana2001}
{Loeb}, A., \& {Barkana}, R. 2001, \araa, 39, 19

\bibitem[{{Mandolesi} {et~al.}(2010){Mandolesi}, {Bersanelli}, {Butler},
  {Artal}, {Baccigalupi}, {Balbi}, {Banday}, {Barreiro}, {Bartelmann},
  {Bennett}, {Bhandari}, {Bonaldi}, {Borrill}, {Bremer}, {Burigana}, {Bowman},
  {Cabella}, {Cantalupo}, {Cappellini}, {Courvoisier}, {Crone}, {Cuttaia},
  {Danese}, {D'Arcangelo}, {Davies}, {Davis}, {de Angelis}, {de Gasperis}, {de
  Rosa}, {de Troia}, {de Zotti}, {Dick}, {Dickinson}, {Diego}, {Donzelli},
  {D{\"o}rl}, {Dupac}, {En{\ss}lin}, {Eriksen}, {Falvella}, {Finelli},
  {Frailis}, {Franceschi}, {Gaier}, {Galeotta}, {Gasparo}, {Giardino}, {Gomez},
  {Gonzalez-Nuevo}, {G{\'o}rski}, {Gregorio}, {Gruppuso}, {Hansen}, {Hell},
  {Herranz}, {Herreros}, {Hildebrandt}, {Hovest}, {Hoyland}, {Huffenberger},
  {Janssen}, {Jaffe}, {Keih{\"a}nen}, {Keskitalo}, {Kisner}, {Kurki-Suonio},
  {L{\"a}hteenm{\"a}ki}, {Lawrence}, {Leach}, {Leahy}, {Leonardi}, {Levin},
  {Lilje}, {L{\'o}pez-Caniego}, {Lowe}, {Lubin}, {Maino}, {Malaspina}, {Maris},
  {Marti-Canales}, {Martinez-Gonzalez}, {Massardi}, {Matarrese}, {Matthai},
  {Meinhold}, {Melchiorri}, {Mendes}, {Mennella}, {Morgante}, {Morigi},
  {Morisset}, {Moss}, {Nash}, {Natoli}, {Nesti}, {Paine}, {Partridge},
  {Pasian}, {Passvogel}, {Pearson}, {P{\'e}rez-Cuevas}, {Perrotta}, {Polenta},
  {Popa}, {Poutanen}, {Prezeau}, {Prina}, {Rachen}, {Rebolo}, {Reinecke},
  {Ricciardi}, {Riller}, {Rocha}, {Roddis}, {Rohlfs}, {Rubi{\~n}o-Martin},
  {Salerno}, {Sandri}, {Scott}, {Seiffert}, {Silk}, {Simonetto}, {Smoot},
  {Sozzi}, {Sternberg}, {Stivoli}, {Stringhetti}, {Tauber}, {Terenzi},
  {Tomasi}, {Tuovinen}, {T{\"u}rler}, {Valenziano}, {Varis}, {Vielva}, {Villa},
  {Vittorio}, {Wade}, {White}, {White}, {Wilkinson}, {Zacchei}, \&
  {Zonca}}]{planck_lfi}
{Mandolesi}, N. {et~al.} 2010, \aap, 520, A3

\bibitem[{{McQuinn} {et~al.}(2006){McQuinn}, {Furlanetto}, {Hernquist}, {Zahn},
  \& {Zaldarriaga}}]{mcquinn_etal2006}
{McQuinn}, M., {Furlanetto}, S.~R., {Hernquist}, L., {Zahn}, O., \&
  {Zaldarriaga}, M. 2006, \nar, 50, 84

\bibitem[{{Mortonson} \& {Hu}(2010)}]{mortonson_hu2010}
{Mortonson}, M.~J., \& {Hu}, W. 2010, \prd, 81, 067302

\bibitem[{{Natarajan} \& {Schwarz}(2008)}]{natarajan_schwarz1}
{Natarajan}, A., \& {Schwarz}, D.~J. 2008, \prd, 78, 103524

\bibitem[{{Natarajan} \& {Schwarz}(2009)}]{natarajan_schwarz2}
---. 2009, \prd, 80, 043529

\bibitem[{{Natarajan} \& {Schwarz}(2010)}]{natarajan_schwarz3}
---. 2010, \prd, 81, 123510

\bibitem[{{Niemack} {et~al.}(2010){Niemack}, {Ade}, {Aguirre}, {Barrientos},
  {Beall}, {Bond}, {Britton}, {Cho}, {Das}, {Devlin}, {Dicker}, {Dunkley},
  {D{\"u}nner}, {Fowler}, {Hajian}, {Halpern}, {Hasselfield}, {Hilton},
  {Hilton}, {Hubmayr}, {Hughes}, {Infante}, {Irwin}, {Jarosik}, {Klein},
  {Kosowsky}, {Marriage}, {McMahon}, {Menanteau}, {Moodley}, {Nibarger},
  {Nolta}, {Page}, {Partridge}, {Reese}, {Sievers}, {Spergel}, {Staggs},
  {Thornton}, {Tucker}, {Wollack}, \& {Yoon}}]{actpol}
{Niemack}, M.~D. {et~al.} 2010, in Society of Photo-Optical Instrumentation
  Engineers (SPIE) Conference Series, Vol. 7741, Society of Photo-Optical
  Instrumentation Engineers (SPIE) Conference Series

\bibitem[{{Pentericci} {et~al.}(2002){Pentericci}, {Fan}, {Rix}, {Strauss},
  {Narayanan}, {Richards}, {Schneider}, {Krolik}, {Heckman}, {Brinkmann},
  {Lamb}, \& {Szokoly}}]{pentericci_etal2002}
{Pentericci}, L. {et~al.} 2002, \aj, 123, 2151

\bibitem[{{Reichardt} {et~al.}(2012){Reichardt}, {Shaw}, {Zahn}, {Aird},
  {Benson}, {Bleem}, {Carlstrom}, {Chang}, {Cho}, {Crawford}, {Crites}, {de
  Haan}, {Dobbs}, {Dudley}, {George}, {Halverson}, {Holder}, {Holzapfel},
  {Hoover}, {Hou}, {Hrubes}, {Joy}, {Keisler}, {Knox}, {Lee}, {Leitch},
  {Lueker}, {Luong-Van}, {McMahon}, {Mehl}, {Meyer}, {Millea}, {Mohr},
  {Montroy}, {Natoli}, {Padin}, {Plagge}, {Pryke}, {Ruhl}, {Schaffer},
  {Shirokoff}, {Spieler}, {Staniszewski}, {Stark}, {Story}, {van Engelen},
  {Vanderlinde}, {Vieira}, \& {Williamson}}]{spt2}
{Reichardt}, C.~L. {et~al.} 2012, \apj, 755, 70

\bibitem[{{Sehgal} {et~al.}(2010){Sehgal}, {Bode}, {Das},
  {Hernandez-Monteagudo}, {Huffenberger}, {Lin}, {Ostriker}, \&
  {Trac}}]{sehgal_etal2010}
{Sehgal}, N., {Bode}, P., {Das}, S., {Hernandez-Monteagudo}, C.,
  {Huffenberger}, K., {Lin}, Y.-T., {Ostriker}, J.~P., \& {Trac}, H. 2010,
  \apj, 709, 920

\bibitem[{{Sokasian} {et~al.}(2003){Sokasian}, {Abel}, {Hernquist}, \&
  {Springel}}]{sokasian_etal2003}
{Sokasian}, A., {Abel}, T., {Hernquist}, L., \& {Springel}, V. 2003, \mnras,
  344, 607

\bibitem[{{Story} {et~al.}(2012){Story}, {Reichardt}, {Hou}, {Keisler}, {Aird},
  {Benson}, {Bleem}, {Carlstrom}, {Chang}, {Cho}, {Crawford}, {Crites}, {de
  Haan}, {Dobbs}, {Dudley}, {Follin}, {George}, {Halverson}, {Holder},
  {Holzapfel}, {Hoover}, {Hrubes}, {Joy}, {Knox}, {Lee}, {Leitch}, {Lueker},
  {Luong-Van}, {McMahon}, {Mehl}, {Meyer}, {Millea}, {Mohr}, {Montroy},
  {Padin}, {Plagge}, {Pryke}, {Ruhl}, {Sayre}, {Schaffer}, {Shaw}, {Shirokoff},
  {Spieler}, {Staniszewski}, {Stark}, {van Engelen}, {Vanderlinde}, {Vieira},
  {Williamson}, \& {Zahn}}]{story_etal}
{Story}, K.~T. {et~al.} 2012, arXiv:1210.7231

\bibitem[{{Su} {et~al.}(2011){Su}, {Yadav}, {McQuinn}, {Yoo}, \&
  {Zaldarriaga}}]{su_yadav}
{Su}, M., {Yadav}, A.~P.~S., {McQuinn}, M., {Yoo}, J., \& {Zaldarriaga}, M.
  2011, arXiv:1106.4313

\bibitem[{{Trac} {et~al.}(2011){Trac}, {Bode}, \& {Ostriker}}]{trac_etal2011}
{Trac}, H., {Bode}, P., \& {Ostriker}, J.~P. 2011, \apj, 727, 94

\bibitem[{{Trac} {et~al.}(2008){Trac}, {Cen}, \& {Loeb}}]{trac_cen_loeb2008}
{Trac}, H., {Cen}, R., \& {Loeb}, A. 2008, \apjl, 689, L81

\bibitem[{{Tumlinson} \& {Shull}(2000)}]{tumlinson_shull}
{Tumlinson}, J., \& {Shull}, J.~M. 2000, \apjl, 528, L65

\bibitem[{{Visbal} \& {Loeb}(2012)}]{visbal_loeb2012}
{Visbal}, E., \& {Loeb}, A. 2012, \jcap, 5, 7

\bibitem[{{Weller}(1999)}]{weller1999}
{Weller}, J. 1999, \apjl, 527, L1

\bibitem[{{Wyithe} \& {Loeb}(2003)}]{wyithe_loeb2003}
{Wyithe}, J.~S.~B., \& {Loeb}, A. 2003, \apjl, 588, L69

\bibitem[{{Zahn} {et~al.}(2012){Zahn}, {Reichardt}, {Shaw}, {Lidz}, {Aird},
  {Benson}, {Bleem}, {Carlstrom}, {Chang}, {Cho}, {Crawford}, {Crites}, {de
  Haan}, {Dobbs}, {Dor{\'e}}, {Dudley}, {George}, {Halverson}, {Holder},
  {Holzapfel}, {Hoover}, {Hou}, {Hrubes}, {Joy}, {Keisler}, {Knox}, {Lee},
  {Leitch}, {Lueker}, {Luong-Van}, {McMahon}, {Mehl}, {Meyer}, {Millea},
  {Mohr}, {Montroy}, {Natoli}, {Padin}, {Plagge}, {Pryke}, {Ruhl}, {Schaffer},
  {Shirokoff}, {Spieler}, {Staniszewski}, {Stark}, {Story}, {van Engelen},
  {Vanderlinde}, {Vieira}, \& {Williamson}}]{spt1}
{Zahn}, O. {et~al.} 2012, \apj, 756, 65

\bibitem[{{Zahn} {et~al.}(2005){Zahn}, {Zaldarriaga}, {Hernquist}, \&
  {McQuinn}}]{zahn_etal2005}
{Zahn}, O., {Zaldarriaga}, M., {Hernquist}, L., \& {McQuinn}, M. 2005, \apj,
  630, 657

\end{thebibliography}
\bibliographystyle{apj}
\end{document}